\documentclass[amssymb,pra,twocolumn,notitlepage,longbibliography,superscriptaddress]{revtex4-1}

\newcommand{\kets}[1]{| #1 \rangle}
\newcommand{\bras}[1]{\langle #1|}
\newcommand{\ket}[1]{| #1 \rangle}
\newcommand{\bra}[1]{\langle #1|}
\newcommand{\bd}[1]{\boldsymbol{#1}}

\usepackage{CJK}
\usepackage[utf8]{inputenc}
\usepackage[T1]{fontenc}
\usepackage{color}

\expandafter\let\csname equation*\endcsname\relax
\expandafter\let\csname endequation*\endcsname\relax
\usepackage{amsmath}                          % better math control.
\usepackage{amsfonts}                         % extended math fonts.
\usepackage{amssymb}                          % names amsfont symbols.
\usepackage{amsthm}                           % theorem environments
\usepackage{mathdots}                         % for ddots going upwards

\usepackage{graphicx}                         % ps, pdf figs.
\usepackage{subfigure}                        % for subfigures
\usepackage{overpic}

\begin{document}

%%%%%%%%%%%%%%%%%%%%%%%%%%%%%%%%%%%%%%%%%%%%%%%%%%%%%%%%%%%%%%%%%%%%%%%%%%%%%%%

\title{Selected topics of quantum computing for nuclear physics}

%\author{Dan-Bo Zhang and Qu-SCNU group(Hui Yan, Shi-Liang Zhu, Hong-Xi Xing, En-Ke Wang)}
%\email{dbzhang@m.scnu.edu.cn}
%\affiliation{Guangdong Provincial Key Laboratory of Quantum Engineering and Quantum Materials, GPETR Center for Quantum Precision Measurement, SPTE, South China Normal University, Guangzhou 510006, China}

%\author{Dan-Bo Zhang$^{1,2}$, Hongxi Xing$^{3,4}$, Hui Yan$^{1,2}$,  Enke Wang$^{3,4}$, and Shi-Liang Zhu$^{1,2}$} %~(QuNu Collaboration)}
%
%\address{$^1$ Guangdong Provincial Key Laboratory of Quantum Engineering and Quantum Materials, School of Physics and Telecommunication Engineering, South China Normal University, Guangzhou 510006, China}
%\address{$^2$ Guangdong-Hong Kong Joint Laboratory of Quantum Matter, Frontier Research Institute for Physics, South China Normal University, Guangzhou 510006,
%China}
%\address{$^3$ Guangdong Provincial Key Laboratory of Nuclear Science, Institute of Quantum Matter, South China Normal University, Guangzhou 510006, China}
%\address{$^4$ Guangdong-Hong Kong Joint Laboratory of Quantum Matter, South China Normal University, Guangzhou 510006, China}
%\address{$^*$ Author to whom any correspondence should be addressed.}
%\eads{\mailto{}, \mailto{}}
\author{Dan-Bo Zhang}
\affiliation{Guangdong Provincial Key Laboratory of Quantum Engineering and Quantum Materials, School of Physics and Telecommunication Engineering, South China Normal University, Guangzhou 510006, China}
\affiliation{Guangdong-Hong Kong Joint Laboratory of Quantum Matter, Frontier Research Institute for Physics, South China Normal University, Guangzhou 510006,
China}

\author{Hongxi Xing}
\affiliation{Guangdong Provincial Key Laboratory of Nuclear Science, Institute of Quantum Matter, South China Normal University, Guangzhou 510006, China}
\affiliation{Guangdong-Hong Kong Joint Laboratory of Quantum Matter, South China Normal University, Guangzhou 510006, China}

\author{Hui Yan}
\affiliation{Guangdong Provincial Key Laboratory of Quantum Engineering and Quantum Materials, School of Physics and Telecommunication Engineering, South China Normal University, Guangzhou 510006, China}
\affiliation{Guangdong-Hong Kong Joint Laboratory of Quantum Matter, Frontier Research Institute for Physics, South China Normal University, Guangzhou 510006,
	China}
\author{Enke Wang}
\affiliation{Guangdong Provincial Key Laboratory of Nuclear Science, Institute of Quantum Matter, South China Normal University, Guangzhou 510006, China}
\affiliation{Guangdong-Hong Kong Joint Laboratory of Quantum Matter, South China Normal University, Guangzhou 510006, China}

\author{Shi-Liang Zhu}
\affiliation{Guangdong Provincial Key Laboratory of Quantum Engineering and Quantum Materials, School of Physics and Telecommunication Engineering, South China Normal University, Guangzhou 510006, China}
\affiliation{Guangdong-Hong Kong Joint Laboratory of Quantum Matter, Frontier Research Institute for Physics, South China Normal University, Guangzhou 510006,
	China}

\date{\today}

\begin{abstract}
Nuclear physics, whose underling theory is described by quantum gauge field coupled with matter, is fundamentally important and yet is formidably challenge for simulation with classical computers. Quantum computing provides a perhaps transformative approach for studying and understanding nuclear physics.  With rapid scaling-up of quantum processors as well as advances on quantum algorithms, the digital quantum simulation approach for simulating quantum gauge fields and nuclear physics has gained lots of attentions. In this review, we aim to summarize recent efforts on solving nuclear physics with quantum computers. We first discuss a formulation of nuclear physics in the language of quantum computing. In particular, we review how quantum gauge fields~(both Abelian and non-Abelian) and its coupling to matter field can be mapped and studied on a quantum computer.  We then introduce related quantum algorithms for solving static properties and real-time evolution for quantum systems, and show their applications for a broad range of problems in nuclear physics, including simulation of lattice gauge field, solving nucleon and nuclear structure,  quantum advantage for simulating scattering in quantum field theory, non-equilibrium dynamics, and so on.  Finally,  a short outlook on future work is given.
\end{abstract}

%\noindent{\it Keywords}: quantum computing, nuclear physics, quantum field theory, quantum simulation, quantum algorithm

\maketitle

%\ioptwocol

\section{Introduction}
Understanding how elementary particles form nuclear matter is  fundamentally important. While the underling basic physical theory, namely quantum chromodynamics~(QCD), can be formulated concisely, it is notoriously hard to solve~\cite{Peskin1995AnIT}. This is because interactions between quarks and gluons follow very distinct behavior: quarks are asymptotically free at the high-energy scale, while strong couplings exist among quarks and gluons at low-energy. In this regard, QCD can be non-perturbative, leading to a failure of the perturbative expansion of Feynman diagrams which have gained remarkable success in quantum electrodynamics~(QED). Lattice QCD has been developed~\cite{wilson_74,kogut_75,kogut_83}, which can rely on computational methods such as quantum Monto Carlo~\cite{mclerran_81,troyer_05} and tensor networks ~\cite{silvi_14,tagliacozzo_14,rico_14,kuhn_15,buyens_16, banuls_17,silvi_19,patrick_20}. The former suffers from the fermion sign problem. The tensor network approach expresses many-body wavefunction in a compressed way which is free of sign problem, but it may have an exponential growth of complexity for evaluating physical observations. While those numeral efforts are still pushing the limits on what quantum many-body problems can be solved on classical computers, they will  meet some intrinsic difficulties that is guaranteed by computational complexity, which claims quantum many-body problems are NP-hard~\cite{troyer_05}.

To overcome the intrinsic difficulty of simulating a quantum world with classical computers, Feynman proposed quantum hardware as simulation platforms in 1983~\cite{feynman_82}. Since then, simulation of physical systems has been an incentive for building quantum computers. Basically, quantum simulation can be divided into simulating static properties and real-time evolution of a physical system~\cite{lloyd_96}. The corresponding quantum algorithms have been developed at the early age of quantum computing based on ideal  quantum computers. Recent scaling-up of quantum processors enables us to simulate static properties and real-time evolution of a quantum system to a larger size with noisy qubits. In such near-term noisy intermediate quantum(NISQ) era~\cite{preskill_quantum_2018}, quantum algorithms are designed suitable for near-term quantum devices. Variational quantum algorithms are remarkable as candidates for fully exploiting the power of NISQ quantum computers. It receives special interests as it provides a practical approaches for solving quantum chemistry~\cite{yung_14,mcclean_16,omalley_16,kandala_17,grimsley_19,arute2020hartreefock}, quantum many-body systems~\cite{liu_19,kokail_19,dallairedemers2020application}, and many other applications~\cite{anschuetz_18,xu2019variational,lubasch_20}.

Nuclear physics can be considered as quantum many-body physics on and below the subatomic scale, and it is of great advantages in using quantum computers for ultimate solutions. However, simulation of nuclear physics meet some special challenges that are different than those of quantum chemistry and quantum many-body physics in condensed matter. Firstly, nuclear physics is studied via  quantum field  theory on continuous space-time, possible with infinite dimensional degrees of freedom per volume, and they should be encoded efficiently into an finite number of qubits on lattice. Moreover, the underling theory for nuclear physics should respect gauge-invariance, and in the corresponding quantum simulation, a physical Hilbert space should be kept. Those two challenges not only call for a formulation of lattice gauge theory for the purpose of efficient quantum computing~\cite{nu_19,zohar_17-frAQB}, but also the issue of realization of reliable simulation on noisy quantum processors becomes tremendous important~\cite{nu_19,stryker_19,halimeh_reliability_2020}.

Efforts of simulating nuclear physics often benefit from advances on quantum simulations of condensed matter or electronic structures of molecules. There are different approaches for tackling this problem, and we may loosely divide it into two types: analog and digital quantum simulation, which may be developed with different motivations. Analog quantum simulation stands for early attempts for simulating gauge lattice theory, when people have realized that cold atoms can simulate quantum many-body problems in condensed matter. Different platforms for quantum simulation of lattice gauge field have been proposed, including cold atoms~\cite{buchler_05,lewenstein_07,boada_10,bermudez_10,banerjee_12,zohar_12,zohar_13,zohar_13-rQF69,zohar_15,stannigel_14,gorg_19,mil_20}, trapped ions~\cite{blatt_12,hauke_13,martinez_16,clark_18,kokail_19,davoudi_20}, superconducting circuit~\cite{marcos_13,marcos_14}, and Rydberg atoms~\cite{tagliacozzo_13,tagliacozzo_13-AzBla,zhang_18,bernien_17,surace_20}. Existing reviews can refer to Ref.~\cite{wiese_13} and Ref.~\cite{zohar_15}  for details, and related artificial gauge fields for ultracold atoms were reviewed in Ref.~\cite{DWZhang2018,DWZhang2011}.
Those proposals stress on how to realize dynamical gauge field with controllable atomic interactions. It is realized all proposals would inevitably require very complicated setup, suggesting that simulation of gauge field and nuclear physics is in general difficult and rather resource consuming. To date, a building block for simulating a coupling of gauge-field with fermion has been implemented in cold-atom experiment~\cite{mil_20}, paving the way for realistic analog quantum simulation of gauge lattice theory.

The approach of digital quantum simulation for gauge field theory and nuclear physics begins later but receives more attention recently~\cite{jordan_12,notarnicola_15,lamm_18,klco_18,nu_19,nu_20,klco_19,klco_20,martinez_16,kokail_19,yeter-aydeniz_19,zohar_17,zohar_17-frAQB,zohar_18,bauer_quantum_2019,raychowdhury_20, du_quantum_2020,roggero_quantum_2020,kreshchuk_light-front_2020,de_jong_quantum_2020,Davoudi_2020yln,wei_quantum_2020,klco_fixed-point_2020,holland_optimal_2020,avkhadiev_accelerating_2020,kharzeev_real-time_2020,shaw_quantum_2020,liu_quantum_2020,mueller_deeply_2020,bepari_towards_2020,kreshchuk_light-front_2020,kreshchuk_quantum_2020}. The digital way is programmable by compiling all operators into basic quantum gates and thus is much more flexible for simulating different quantum systems~\cite{nielsen_chuang_2010}. In 2012, Preskill \emph{et al} proves an exponential quantum speed-up in simulation of scattering problem for scalar relativistic quantum field theory with self-interactions~($\phi^4$ theory)~\cite{jordan_12} , with time complexity scaling up polynomial in the number of particles, their energy and the desired precision, remarkably in the non-perturbative regime where classical algorithms fail to work. Although elaborated on a concrete model, it puts a solid base for  simulating high-energy physics with quantum advantages from the aspect of computational complexity.
More quantum algorithms have proposed to simulate lattice gauge field and nuclear physics on near-term quantum processors. While lots techniques can inherit from those of quantum computational chemistry, one key difference is the requirement to deal with gauge field and gauge-invariance. A very successful example is that gauge degree of freedom for the 1+1D Schwinger model can be eliminated~\cite{Muschik_2017,martinez_16,kokail_19}. With this reduction, simulations of ground state and real-time evolution of 1+1D Schwinger model have been demonstrated on trapped-ion platforms~\cite{martinez_16,kokail_19}. However, in general such an elimination is impossible, and systematic formulations of lattice gauge theory suitable on quantum computers have been investigated, especially on how to reduce the desired quantum resource for gauge field which has much redundancy. The approach from lattice gauge theory, however, is still too resource demanding for simulating nuclear physics interested in experimental observations. Another practical strategy, mostly pushed by the community of high-energy and nuclear physics, is to incorporate quantum algorithms as subroutines and wisdom of classical methods are exploited. This permits valuable quantum resources concentrated on classical difficulty parts, in order to exploit near-term noisy quantum processors to solve complicated nuclear physics. Such a hybrid quantum-classical strategy has been used in studying parton distributions~\cite{nu_20}, evolution of non-equilibrium thermal states such as quark-gluon plasma~\cite{lamm_18}, and so on.

Although sufficient advances have been made, quantum simulation of high-energy nuclear physics is still a young field, and it is just beginning to bring more researchers into this field. In this review, we aim to summarize recent advances, give basic concepts on quantum computing and quantum algorithms, introduce several representative works, and propose future directions. The review will be organized as follows.

In Sec.~\ref{sec:nuclear2qc}, we first discuss how nuclear physics problems can be mapped into formulas of lattice field theory expressed with qubits, which can be solved on a quantum computer. Specifically, we reveal that the way of treating gauge field may be the key ingredient. We then introduce quantum algorithms related for simulating quantum systems, including both static properties and real-time evolution.

In Sec.~\ref{sec:applications}, we give some specific topics and representative examples on the applications of quantum computing for nuclear physics. We first elaborate on a prototypical example, the 1+1D Schwinger model. This model, although describing quantum electrodynamics~(QED), shares lots of key features with QCD, such as confinement and a topological theta vacuum. Thus, this minimal model can be used as a testbed for simulating lattice gauge field.  We also give examples on hybrid quantum-classical approach which uses quantum algorithms as a subroutine, including examples of parton physics and evolution of non-equilibrium thermal states.
% While the above example focus on digital quantum simulators, we also present how analog quantum simulator can simulate gauge quantum field and the related nuclear physics.

Finally, in Sec.~\ref{sec:outlook}, we give outlooks and summaries.

\section{Nuclear physics on a quantum computer}\label{sec:nuclear2qc}
In this section, we first discuss how to map nuclear physics problems on to a quantum computer, by reformulating gauge quantum field in the language of qubits. Then, we introduce quantum algorithms for solving static and dynamical properties of nuclear physics.

\subsection{Map nuclear physics onto a quantum computer}
The underling theory for nuclear physics is quantum field theory describing fermionic matter coupled with bosonic gauge fields on a continuous space-time background. However, a quantum computer consists of arrays of qubits, and each qubit owns 2-dim Hilbert space.
%  and is bosonic in nature.
To simulate nuclear physics on a quantum computer, it demands for a mapping of the original physical degree of freedom onto qubits at the first stage.

\begin{figure*}[htbp] \centering
	\includegraphics[width=15cm]{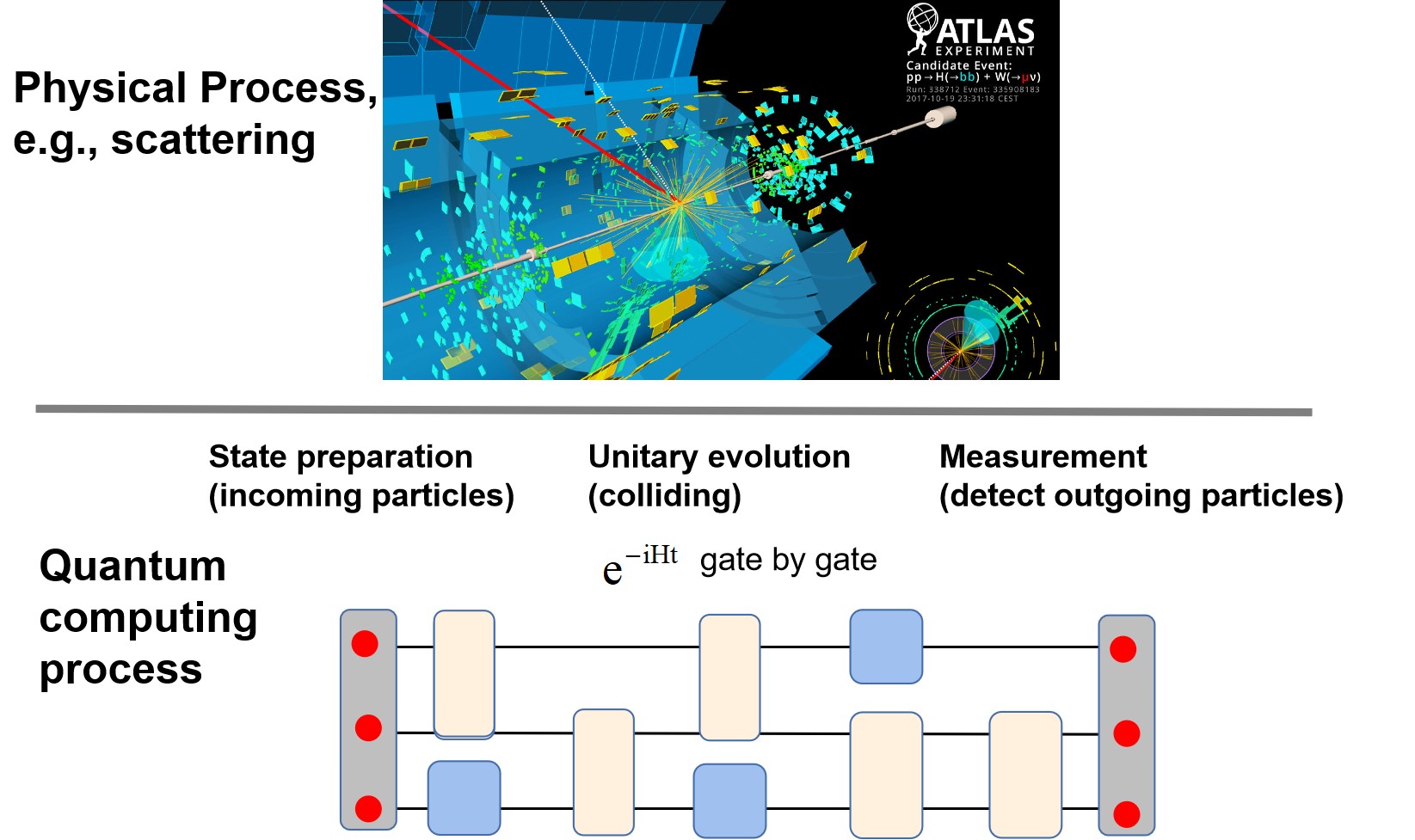}
	\caption{Illustration of solving nuclear physics on a quantum computer. The upper is a physical process of scattering in the experiment of Large Hadron Collide~(from https://www.symmetrymagazine.org/article/lhc-scientists-detect-most-favored-higgs-decay), while the bottom depicts a  corresponding simulation with a quantum computer.}
	\label{fig:nucleartoquantum}
\end{figure*}

On a quantum computer, the basic unit to encode information is quantum bit, which can store a superposed state of $\ket{0}$ and $\ket{1}$. In general, information is expressed as a quantum state of multiple qubtis.  With several hundreds of qubits, a quantum computer can store a quantum state that is beyond the capacity of classical computers. The quantum state is manipulated on a quantum computer with unitary evolutions, which can be decoupled into a sequence of quantum gates from a universal set of basic quantum gates. Then, information or interested quantities can be extracted by repeatedly measuring the final quantum states. Further post-processing on a classical computer may be required.  The physical process and its corresponding simulation on a quantum computer is illustrated in Fig.~\ref{fig:nucleartoquantum}.

Nuclear physics includes problems related to steady states, such as structure of nucleons and phases of matter, and  real-time evolution, such as scattering problem and evolution of nuclear matter. While preparation of steady states and simulating the real-time evolution have been standard techniques in quantum computing and variants of approaches have been developed, it is important to firstly convert nuclear physics into formulas that a quantum computer can handle with. Keep in mind that basic components of a quantum computer are qubits for encoding information, universal set of single and two-qubit quantum gates for manipulating information, and measurements for extracting information.  To study nuclear physics on a quantum computer, those field operators should be converted into qubit operators, and the space-time should be discretized, and some symmetries should be considered, especially the local gauge invariance~(general Gauss law). We discuss them separately.

\textbf{Encode fields into qubits on lattice.} Quantum fields usually are described on a continuous space-time and they can be bosonic~(infinite dimension) or fermionic, while qubits on a quantum computer is defined on a lattice with each site resides one qubit. To encode quantum state of a quantum field on a quantum computer, we should discretize the space as a lattice, encode each local bosonic field into a finite number of qubits by a truncation, and encode a local fermionic field into a sequence of qubits by nonlocal transformations. Remarkably, to reduce the computational resource yet retain the essential physics, the discretization of space and the cut-off bosonic field should respect some physical arguments. In other words, the original nuclear physics problem may have an infinite degree of freedoms and it is important to map it into a task solvable on a quantum computer with a finite number of qubits under desired precision~\cite{jordan_12,klco_19}.
Another important ingredient in the encoding stage is how to express gauge field with qubits. It is important to enforce local gauge-invariance~(general Gauss law) for simulating lattice gauge theory on a quantum computer~\cite{zohar_17-frAQB,nu_19,stryker_19,patrick_20} which corresponds to keep the simulation in the physical Hilbert space.

\textbf{Unitary evolution.} Physical processes in nuclear physics can be both unitary and nonunitary, while states on a quantum computer evolve in a unitary process. For an unitary evolution such as evolutions of the Hamiltonian, one can decompose the unitary operator into a sequence of basic single and two-qubit quantum gates with product-formula, by cutting the evolution time into short time periods.  As an nonunitary process can be embedded in an unitary process of a larger system, one can introduce some auxiliary qubits to implement the nonunitary evolution. A common applied algorithmic design is by linear-combination-of-unitaries, using ancillary qubits ~\cite{gui-lu_06,childs_17,van_17,chowdhury_17,gilyen_18,arrazola_19} or continuous-variables~\cite{lau_17,arrazola_19,zhang_19,zhang_20,zhang2020continuousvariable}.

%It is also noted that unitary process on a quantum computer may also be constructed with Hamiltonian evolutions, which is referred as analog quantum computer or simulator.
%The merit of analog approach is that

\textbf{Measurements.}	
In nuclear physics, we often need to get distributions of particles, such as in the scattering problems. On a quantum computer, however, measurement often refer to computational basic ($\sigma^z$ measurements), which may not directly correspond to find particles in specified directions or momentums. One can synthesize  such physical measurements by unitary evolutions followed by measurements on computational basis.

%textbf{Symmetry considerations.}
%Most importantly, quantum fields have special requirements, such as gauge invariant and Lorentz invariant. It is important to consider how to handle those intrinsic symmetry considerations on a quantum computer. Unfortunately, while discretized a quantum field and formulated in a Hamiltonian formula, the Lorentz symmetry also loses, and it calls for a fine tuning of the bare Hamiltonian to describe continuum physics of the quantum field.

\textbf{Digital \emph{vs} analog quantum computer.}
It should be reminded of a difference between digital and analog quantum computers.
Analog quantum computers are designed specifically for simulating quantum systems. Unlike a digital quantum computer that uses an universal set of quantum gates to construct all unitary process, an analog quantum computer can directly simulate a Hamiltonian and its real-time evolution, by engineering the desired interactions. This limits its capacity for simulating different systems, while on the other hand, it can be easier to implement. Representative physical systems for analog quantum computers include cold atoms ~\cite{DWZhang2018}, trapped ions, and superconducting circuits. Nevertheless,  analog quantum simulator can be programmable, enlarging its capacity of simulating a different target Hamiltonian with a controllable Hamiltonian on the physical platform using variational methods. For instance, ground state of the 1+1D Schwinger model has been simulated in the trapped-ion system with long-interaction between qubits~\cite{kokail_19}.

\subsection{Quantum algorithms for simulation}
\label{sec:quantum_algorithm}
To exploit the power of quantum computers for simulating nuclear physics, quantum algorithms are indispensable.  After having converting a nuclear physics problem on a quantum computer, the remaining task is to design a quantum algorithm to solve it efficiently. Although quantum algorithms can be counter-intuitive and requires special efforts to learn in general, it can be very natural  for physicists when using  quantum algorithms for solving quantum systems. Those quantum algorithms  may be classified as two kinds: preparing steady states of a quantum system and simulating the real-time evolution. In the following, we give a brief introduction.

Simulating Hamiltonian evolutions $e^{-iHt}$ is typically hard for classical computers, due to the growth of complexity of quantum states with increasing $t$. However, it can be very directly implemented on a quantum computer, by decomposing $e^{-iHt}$ into a set of quantum gates~\cite{lloyd_96}. For a Hamiltonian $H=\sum_{i=1}^M H_i$ with local terms $H_i$ that $[H_i,H_j]=0$ for $i\neq j$, the most basic technique is to use a product formula~(~also named as Trotter formula), which is
\begin{equation}\label{eq:trotter_evolution}
e^{-iHt} \approx\left[\prod_{i=1}^{M}e^{-iH_i\frac{t}{N}}\right]^N.
\end{equation}
For an desired precision $\epsilon$, the time complexity is $O(t^2/\epsilon)$.
It is interesting to note that while such a formula is simple, it is powerful even on the near-term quantum devices, and the performance can be comparable with more advanced quantum algorithms for Hamiltonian simulations~\cite{low_17,campbell_19,childs2019theory}. Remarkably, it is found that time-complexity can be reduced to $O(t/\epsilon)$ using the product-formula when the Hamiltonian can be written as $H=H_A+H_B$, where local terms in $H_A$ or $H_B$ commute to each other but $[H_A,H_B]\neq 0$. This is because there is a destructive error interference that errors will cancel in different short time periods~\cite{tran_20}. This indicates product-formula may be more efficient than expected.

Solving steady states, including eigenstates and thermal states, is important for understanding static properties of quantum systems, but can be formidable challenge on a classical computer. Such a basic task is comparatively harder on a quantum computer than simulation of real-time evolution, and lots of efforts have been devoted into it. The first textbook algorithm is quantum phase estimation~\cite{nielsen_chuang_2010}, where eigenvalues of Hamiltonian are written onto ancillary qubits and then ancillary qubits are measured to read eigenvalue as well as associated eigenstates. Concretely, consider $H\kets{u_n}=E_n\kets{u_n}$ and the initial state is $\ket{\psi}=\sum_n c_n \ket{u_n}$, then each eigenstate can be obtained with a probability $|c_n|^2$. Thus it is important that the target eigenstate should have a large weighting in the initial state.
Quantum phase estimation can be understood as signal processing from the time domain to the frequency domain. It involves Hamiltonian evolution at different time periods, and then uses quantum Fourier transformation to find eigenvalues at frequency domain. Both Hamiltonian evolution and quantum Fourier transformation are resource-costing, and consequently, quantum phase estimation is not suitable on the near-term quantum devices.  Quantum adiabatic algorithm provides an approach to prepare ground state of a Hamiltonian~\cite{albash_18}. It starts from a Hamiltonian whose ground state is easy to prepare, and by adiabatically tuning the Hamiltonian into the target Hamiltonian, the final state will be the ground state for the target Hamiltonian. The performance of the quantum adiabatic algorithm relies on a gap between the first excited state and the ground state along the adiabatic path. Thus it requires a good design for the adiabatic path which can be nontrivial. In practice, the adiabatic evolution should be very slow, and may be hard to finish within the coherent time. Thus,  quantum adiabatic algorithm, although can be universal, may not be suitable in the era of NISQ.
%Nevertheless, it has been applied for solving combinational optimization problems.

Another important class of quantum algorithms is variational quantum eigensolver~(VQE)~\cite{yung_14, mcclean_16,shen_17,kandala_17,hempel_18,liu_19,kokail_19,grimsley_19,danbo_20}, which is considered  promising for fully exploiting the power of NISQ quantum devices. It can rely on a shallow quantum circuit and a moderate number of qubits for solving classical intractable problems. The quantum circuit is parameterized and the parameters can be obtained by minimizing the energy. The optimization is a hybrid quantum-classical one. In this sense, variational quantum algorithms give a new paradigm for algorithmic design: rather than directly designing a quantum algorithm, it trains a quantum algorithm for a given task, by optimizing a cost function.
The VQE is designed for solving the ground state by minimizing the energy, and variants of VQE have been developed for obtaining excited states~\cite{mcclean_17,nakanishi_19,Higgott_19,zhang_variational_2020}, thermal states at finite temperature~\cite{liu_19_gibbs,wu_19,zhu2019generation,chowdhury2020variational,wang_20}, quantum imaginary time evolution~\cite{mcardle_19}, and general quantum processes~\cite{endo_variational_2020}.

\section{Applications for nuclear physics}\label{sec:applications}

Although at an early stage, quantum computing has had a broad applications for nuclear physics problems, and it is beyond the scope of this review for a throughout investigation. Rather,   we attend to use some prototypical examples to illustrate how nuclear physics problems can be solved on a quantum computer. With concrete examples, the basic concepts and procedures may be revealed. We start with the 1+1D Schwinger model, which is a prototypical model for simulating gauge field with a state-of-art realistic quantum computer. Other examples include interesting problems ranging from scattering, evolution of non-equilibrium thermal states, nuclear structure at both high-energy and low-energy, and so on. In addition, we introduce a work that give the time complexity of solving scattering for a scalar quantum field, which shows quantum advantage.

\subsection{Simulation of real-time evolution and ground state of lattice gauge theory: the Schwinger model}
The capacity of quantum simulation of lattice gauge field implies that nuclear physics can be studied on a quantum computer. However, the required quantum resource is too demanding in the NISQ era, and it is more practical to start with some minimal models. The Schwinger model is such a prototypical model that has been demonstrated  on current quantum devices, including both ground state~\cite{kokail_19} and real-time evolution~\cite{martinez_16}.  Moreover, the Schwinger model   can reveal some important features shared with QCD. We thus use this model to illustrate a work flow for studying nuclear physics on a quantum computer: how the original formula of quantum field problem should be mapped into a lattice spin model, what quantum algorithm should be chosen to solve the lattice spin model, and how desired physical quantities can be accessed with measurements on a quantum computer, possible with post-processing.

We introduce the Hamiltonian for the Schwinger model with fixing gauge $A_0(x)=0$,
\begin{eqnarray} \label{eq:Schwinger_density}
\hat{H}=\int dx [\Psi^\dagger(x)\gamma^0\gamma^1(-i\partial_1+g\hat{A}_1(x))\Psi(x) \nonumber \\
+ m\Psi^\dagger(x)\gamma^0\Psi(x)+\frac{1}{2}\hat{E}^2(x)],
\end{eqnarray}
where $\Psi^\dagger(x), \Psi(x)$ are fermionic field operators, the electric field $\hat{E}(x)$ and the vector potential $\hat{A}_1(x)$ satisfy $-\hat{E}(x)=\partial_0\hat{A}_1(x)$ and $[\hat{A}_1(x), \hat{E}(x')]=-i\delta(x-x')$,  $\gamma^0=\sigma^z$ and $\gamma^1=i\sigma^y$. In addition, the local conservation~(Gauss law) should be respected, $\partial_1E=g\Psi^\dagger(x)\Psi(x)$, which is a consequence of local $U(1)$ gauge symmetry.

A recasting of Eq.~\eqref{eq:Schwinger_density} onto a lattice is the Kogut-Susskind formula~\cite{susskind_77}, which puts fermions on the sites and gauge field on bonds. Concretely, two-component fermions are defined on the even and odd sites respectively: $\Phi_{2j}=\sqrt{a}\Psi(x_{2j})$ and $\Phi_{2j-1}=\sqrt{a}\Psi^\dagger(x_{2j-1})$.  Gauge fields on the bonds are $\theta_{j,j+1}=-agA(x_j+\frac{a}{2})$ and $L_{j,j+1}=\frac{1}{g}E(x_j+\frac{a}{2})$, which satisfy $[\hat{\theta}_{j,j+1},\hat{L}_{j',j'+1}]=i\hat{\delta}_{j,j'}$. By those representations, the lattice Hamiltonian under open boundary reads,
\begin{eqnarray} \label{eq:Schwinger_lattice}
\hat{H}&=& \frac{1}{2a}\sum_{j=1}^{N-1}[\Phi_j^\dagger e^{i\hat{\theta}_{j,j+1}} \Phi_{j+1}+h.c.] \nonumber \\
&+&m\sum_{j=1}^{N}(-1)^j\Phi_j^\dagger\Phi_j + \frac{g^2a}{2}\sum_{j=1}^{N-1}\hat{L}^2_{j,j+1}.
\end{eqnarray}

The lattice Hamiltonian has already allowed for numeral simulation. Still, a challenge is to simulate the continuous variable gauge field with discrete-variable qubits. Approximations can be applied, either by reduction the $U(1)$ to $Z_n$ gauge field, or by a truncation into finite-dimension Hilbert space. Those can be standard schemes for all Abelian and non-Abelian gauge fields. Fortunately, the 1+1D Schwinger model is special as there is essentially no magnetic field. Moreover, the electric field and fermions are locally dependent into each other. This allows us to eliminate the gauge field totally by using the Gaussian law, $\hat{L}_{j,j+1}-\hat{L}_{j-1,j}=\Phi_j^\dagger\Phi_j-\frac{1-(-1)^j}{2}$, which determines $\hat{L}_{j,j+1}=\epsilon_0+\sum_{i=1}^{j}[\Phi_i^\dagger\Phi_i-\frac{1-(-1)^i}{2}]$, with $\epsilon_0$ as the value of $\hat{L}$ at the left boundary.  Now, the lattice model can be represented solely with fermionic operators, and by Jordan-Wigner transformation, $\Phi_j^\dagger=\prod_{i<j}\sigma^z_i\sigma_j^+$, it can be written as a qubit Hamiltonian,
\begin{eqnarray} \label{eq:Schwinger_qubit}
\hat{H}= \frac{1}{2a}\sum_{j=1}^{N-1}[\sigma_j^+\sigma_{j+1}^-+\sigma_j^-\sigma_{j+1}^+] +\frac{m}{2}\sum_{j=1}^{N}(-1)^j\sigma_j^z\nonumber \\
 +\frac{g^2a}{2}\sum_{j=1}^{N-1}\left(\epsilon_0-\sum_{i=1}^{j}(\sigma_i^z+(-1)^i)\right)^2.
\end{eqnarray}

By eliminating the gauge field, the Hamiltonian is a bit more complicated as there are long-range interactions, but is still feasible on current quantum processors.  Remarkably, trapped-ion quantum computers are renowned for its excellent connectivity, which is naturally suitable for simulating quantum systems with long-range interactions, while other platforms may suffer from limited connectivity. The following two experiments on trapped ions present the state-of-art simulations for the lattice gauge field on the real-time evolution and ground state properties, respectively.

\textbf{Real-time evolution.}
Many nuclear physics phenomena involve time evolution and it is fundamentally important to simulate the  real-time evolution of gauge theory. While existing classical methods often meet difficulties, simulation of real-time evolution can be implemented very naturally on a quantum computer, once an real-time evolution of a Hamiltonian is compiled into a sequence of quantum gates with Trotter decomposition. The Schwinger model was raised for explaining the mechanism of creating pairs of fermion-antifermion from the bare vacuum. Such a process can be simulated with a language of qubits. Firstly, occupation of a fermion~(antifermion) is encoded as $\ket{1}$($\ket{0}$) on the even~(odd) site. Then, the bare vacuum has no fermion at all, which can be represented as $\ket{0101010101...}$. With time evolution of the Hamiltonian, pairs of fermion-antifermion will appear due to hopping terms in the Hamiltonian. To implement $e^{-itH}$ on a quantum computer, a Trotter formula can decompose the time evolution into a product of short-time evolution of small Hamiltonian term, as in Eq.~\eqref{eq:trotter_evolution}. Proliferation of fermion-antifermion pairs after a period of $t$ can be revealed by repeated measurements.

In trapped ions, a qubit is encoded as two internal energy levels in an ion. The experiment uses four ions to demonstrate the creation of fermion-antifermion pairs on realistic trapped-ion platforms, and the result fits good with theoretical results, with both idea evolution and the case of using the Trotter decomposition. Although with only a few qubits, this work has paved the way for simulating dynamical systems of large-scale systems that basically follows the same scheme.

\textbf{Ground state.}
While real-time evolution for a system can be implemented directly on a quantum computer, solving its ground state or excited states requires more efforts on quantum algorithms.
%Although quantum phase estimation offers an universal scheme, it requires to firstly prepare the system in a quantum state with a large overlapping with the ground state, and the quantum circuit for sampling the ground state and reading the ground state energy has a deep depth, which challenges the current quantum processors.
Variational quantum eigensolver provides a feasible scheme on the NISQ quantum processor.  The key point is to train a quantum circuit to prepare a variational ground state $\ket{\Psi(\boldsymbol{\theta})}=U(\boldsymbol{\theta})\ket{\Psi_0}$ for the target Hamiltonian $H$, where parameters are optimized by minimizing the energy, $E(\boldsymbol{\theta})=\bras{\Psi(\boldsymbol{\theta})}H\ket{\Psi(\boldsymbol{\theta})}$.  For the Schwinger model, one challenge
is to adopt a wavefunction ansatz with enough expressive power under limited quantum resource, e.g., the parametrized quantum circuit should be shallow and yet can describe complicated quantum correlation of the ground state due to the long-range interactions. The current digital quantum computer is limited to small sizes as the required depths of quantum circuit can increase quickly with the system size. On the other hand, analog quantum simulator can realize an unitary operator by directly letting an engineered Hamiltonian evolves, which may be decomposed into a long sequence of quantum gates. Ref.~\cite{kokail_19} devises variational quantum simulator that can simulate the target Hamiltonian with engineered different Hamiltonian.

For simulating the Schwinger model with trapped-ions, an advantage is that engineered Hamiltonian of trapped-ions naturally have long-range interactions, which can be written as,
\begin{eqnarray}
H_R^{(0)}=\sum_{i\neq j}J_{i j}(\sigma_i^+\sigma_j^-+\sigma_i^-\sigma_j^+) + B \sum_j\sigma^z_j,
\end{eqnarray}
where $J_{ij}\simeq J_0/|i-j|^\alpha$, with $0<\alpha<3$ that is tunable    long-range interactions.
Another engineered Hamiltonian required is $H_R^{(j)}=\frac{\Delta}{2}\sigma^z_j$ that generates local rotation along $z$ axis.
Those resource Hamiltonian can be applied to generate a trivial state via  $U(\boldsymbol{\theta})=\exp(-i\theta_kH_R^{(i_k)})...\exp(-i\theta_kH_R^{(i_1})$. An ansatz is to arrange odd layers as $\exp(-i\theta H_R^{(0)})$ and even layers as $\exp(-i\theta H_R^{(j)})$ performing independently on each qubit. The initial state is chosen as the bare vacuum state $\ket{\Psi_0}=\ket{01...01}$, corresponding to the ground state at $m\rightarrow \infty$.  The ansatz always keep the state in the subspace of equal fermion and antifermion, which respects the symmetry of the ground state.

The parameter vector $\boldsymbol{\theta}$ is trained by minimizing the variational energy $E(\boldsymbol{\theta})$.  The optimization can be nontrivial for a noisy quantum processor, and it adopts a global optimizer called DIRECT~\cite{kokail_19}, which divides the parameter space into cells for search.  It is demonstrated that the trapped-ion simulator can variationally solving the ground state for up to 20 qubits for the Schwinger model, which is quite remarkable. Moreover, the experiment use zero energy variance to self-verifying that the obtained state is indeed the eigenstate, since an eigenstate can be characterized by zero energy variance. Although only shown with eight qubits,  self-verifying can be vital for large-size quantum simulator that is beyond the computational capacity of classical computers:  it requires the quantum computer itself to verify if it gives a result with enough accuracy. In addition,  a quantum phase transition is revealed on the trapped-ion simulator by tuning the mass across $m_c\approx-0.7$(set $a=1$ and $g=1$). This indicates that the variational quantum simulator can efficiently solve ground states for the Schwinger model for a large range of parameters, especially around the quantum critical point where quantum correlations is strong.  The work opens a new direction for simulating large-size quantum system exploiting analog quantum simulators with great flexibility by using the variational method.

We comment that elimination of the gauge field completely is only possible for 1+1D $U(1)$ gauge theory. Nevertheless, the idea of reducing quantum resources for representing gauge field is important and possible, since gauge field has a redundancy in description.  Many efforts have been denoted for this mapping stage, including non-Abelian gauge field in 1+1D~\cite{banuls_17,klco_20}, 1+2D Abelian~\cite{zohar_17} and non-Abelian gauge field, and so on .

\subsection{Nuclear structure}
One center motivation of nuclear physics is to explore the internal structure of nucleus, which relies on an increasing of energy scale of probes for resolving finer structures.  Protons and neutrons are basic ingredients of a nucleus at low-energy scales, but a proton or a neutron itself is a collection of quarks and gluons at large energy scales.  Colliders have been built for detecting those structures at different energy scales from scattering cross sections.  Yet, numeral methods meet challenges that are common for quantum many-body systems, especially for solving the structure of hardrons, which in nature is non-perturbative.  We review recent efforts for solving nuclear structure on a quantum computer~\cite{dumitrescu_18, nu_20},  for an atomic nucleus and an hadron respectively, which adopt  quite different strategies.

\subsubsection{Binding energy of atomic nucleus}
From the aspect of low-energy nuclear physics, an atom consists of  interacting  protons and neutrons that are bounded together.
Although lattice QCD can be applied,  a better starting point is to use effective field theory~(EFT) that  protons and neutrons are relevant degree of freedoms.  For light nucleus,  pionless EFT provides systematically  improvable approach for modeling nuclear interaction.  For this bound-state problem, a second-quantization of the Hamiltonian can be obtained by using proper local basis, e.g.,  a common choice can be the harmonic oscillator basis. As a truncation of local basis can be applied, the problem can be solved in a finite-dimensional Hilbert space.  Further, the second-quantization fermionic Hamiltonian can be mapped into a qubit Hamiltonian. Such a procedure is very familiar in quantum chemistry and quantum many-body problems in condensed matter.  And one can immediately recognize that the problem is much like quantum chemistry, and it is expected that an exponential wall prevents classical methods when the system size is becoming large.

An interesting demonstration of quantum computing for atomic nucleus was reported in Ref.~\cite{dumitrescu_18}, which adopts the VQE to solve the binding energy of the deuteron on a cloud quantum computer.  With pionless EFT, the deuteron Hamiltonian with a truncation of $N$ basis reads,
\begin{eqnarray} \label{eq:H_deuteron}
H_N=\sum_{nn'=0}^{N-1}t_{nn'}a_{n'}^\dagger a_n,
\end{eqnarray}
where $a_n^\dagger$($a_n$) create~(annihilate) a deuteron in the harmonic-oscillator s-wave sate $\ket{n}$, and the hopping strength $t_{nn'}=\bra{n'}(T+V)\ket{n}$ can be calculated under the harmonic oscillator basis. As the deuteron is described in EFT via a contact interaction, only nearest hopping exists~(see Ref.~\cite{dumitrescu_18} for detailed values).  By the Jordan-Wigner transformation,  the Hamiltonian with increasing $N$~(up to $N=3$) can be written as,
\begin{eqnarray}
H_2&=&5.906 709 I+ 0.218 291\sigma_0^z - 6.125\sigma_1^z  \nonumber\\ &-&2.143304(\sigma_0^x\sigma_1^x+\sigma_0^y\sigma_1^y) \nonumber\\
H_3&=&H_2+9.625(I-\sigma_2^z ) \nonumber\\ &+& 3.913119(\sigma_1^x\sigma_2^x+\sigma_1^y\sigma_2^y). \nonumber\\
\end{eqnarray}

With the harmonic oscillator basis,  the deuteron ground state will be a superposition state at different basis, which is clear due to the hopping terms.  The VQE approach uses an unitary-coupled-cluster~(UCC) ansatz.
Defining single excitation process, $U_{01}(\theta)=e^{\theta(a_0^\dagger a_1-a_1^\dagger a_0)}=e^{i\theta/2(\sigma_0^x\sigma_1^y-\sigma_0^y\sigma_1^x)}$ and $U_{02}(\eta)=e^{\eta(a_0^\dagger a_2-a_2^\dagger a_0)}=e^{i\eta/2(\sigma_0^x\sigma_1^z\sigma_2^y-\sigma_0^y\sigma_1^z\sigma_2^x)}$, then we can employ ansatz $U(\theta)=U_{01}(\theta)$ for $H_2$,
$U(\eta,\theta)=U_{01}(\theta)U_{02}(\eta)$ for $H_3$, both performing on the initial state $\ket{100}$. Parameters $\eta$ and $\theta$ are optimized by minimizing the energy of $H_2$ and $H_3$ with their corresponding variational wavefunciton, respectively.
Further, the extrapolation to the infinite space should be employed with data obtained at finite size. The numeral results are shown to be very accurate with exact diagonalization.

We point out that solving the Hamiltonian in Eq.~\eqref{eq:H_deuteron} is actually easy by exact diagonalization since it involves a
tridiagonal matrix. Why bother to use a quantum computer?  Moreover, a $N\times N$ matrix is easy to diagonalize on a classical computer for $N\sim100$, but can be challenge for the current quantum devices.   It should be reminded that the VQE approach becomes valuable when solving nuclear structure becomes a many-body problem, and a simple exact diagonalization will face a $M\otimes M$ matrix ($M$ grows exponentially with $N$).  It is expected that a road map for extending the scheme for quantum many-body nuclear physics is required.

\subsubsection{Parton physics}
Solving internal structure of a hadron is quite a different story.   The partonic structure of a hadron depends on the energy scale to see it. For instance, a proton is a bounded state of three valence quarks at low energy scale, and  no individual quark has been observed in experiments due to color confinement. On the other hand, at higher energy scale in deep inelastic scattering~(DIS) with large momentum transfer,  the hadron shall be modeled as a collection of charged point-like constituents, namely parton. The parton contributes to the cross section with momentum $xp^\mu$, where x is the momentum fraction of the proton carried by the parton, and $p^\mu$ is the momentum of the hadron.  Understanding hadronic structure in terms of the parton distribution functions~(PDFs) over $xp^\mu$ and its generalization is a research frontier~\cite{ji_parton_2013,Kang:2013raa,Kang:2014lha,lin_parton_2018}.

The PDFs $f(x)$ serve as an input for explaining scattering cross sections in DIS experiments. However, the parton distribution itself is hard to compute from \textit{ab initio} methods due to its nonperturbative nature.  There are different  approaches from first principle,  such as light-front Hamiltonian and lattice QCD, which still awaits for exascale supercomputer to verify.  Here we introduce a recent work that points out an avenue on a near-term quantum computer.

Ref.~\cite{nu_20} presents two different schemes for calculating parton distribution on a quantum computer:  direct calculation of PDFs using operator formula and by  evaluation of the hadronic tensor.  For illustration purpose two schemes are demonstrated with 1+1D Thirring model.  The lattice Hamiltonian writes,
\begin{eqnarray} \label{eq: thirring_H}
H&=&\sum_{j}\frac{1}{2}(-1)^j[c^\dagger_jc_{j+1}+h.c.]\nonumber \\
&+&m(-1)^jc^\dagger_jc_j -g^2n_jn_{j+1}
\end{eqnarray}
where $n_j=c^\dagger_jc_j$.  This model is not gauge theory and thus is different from QCD. However, evaluating the hadronic tensor will not depend on the gauge field explicitly in lepton-hadron DIS at the leading order, and thus the second scheme is plausible and promising for studying PDF on NISQ quantum processors. The Hamiltonian shall be mapped into a spin model,
\begin{eqnarray}
H&=&\sum_{j=1}^{N-1} \frac{(-1)^{j+1}}{4}(\sigma_j^x\sigma_{j+1}^x+\sigma_j^y\sigma_{j+1}^y) \nonumber \\
&+& \sum_{j=1}^{N}(\frac{(-1)^jm}{2}\sigma_j^z+\frac{g^2}{4}\sigma_j^z\sigma_{j+1}^z).
\end{eqnarray}
Similar to the Schwinger model, the bare vacuum is set at $m\rightarrow \infty$ limit, and  $\ket{1}$($\ket{0}$) on the even~(odd) site represents occupation of a fermion~(antifermion), respectively.

The operator formula for a definition of PDF~(without gauge link) is given by,
\begin{equation}\label{eq:operator_pdf}
f(x)=\int dy e^{ixP^+y}\bra{P}\bar{\psi}(y)\gamma^+\psi(0)\ket{P},
\end{equation}
where $\gamma^+=(\gamma^0+\gamma^1)/\sqrt{2}$ and $P^+=(P^0+P^1)/\sqrt{2}$ is the light-cone momentum for the incoming hadron.  Noted that there is no Wilson line  in the definition of Eq.~\eqref{eq:operator_pdf} as gauge field is not involved in the Thirring model in Eq.~\eqref{eq: thirring_H}.
$\ket{P}$  stands for the hadron state at momentum $P$. In this sense, PDF is the Fourier transform of a time-separated correlator of the hadron. Two key procedures on a quantum computer are clear: firstly prepare $\ket{P}$ which is a quantum many-body state for the Thirring model with quantum numbers for representing
the hadron at momentum $P$; then calculate the correlator.  Since there is only one flavor of fermion, we may express a hadron with a quantum number of three fermions, plus fluctuations of fermion-antifermion pairs.  Such a scheme seems very straightforward, but some technical issues prevents it from being useful.  Since light cone can not be defined on the lattice until the continuum limit,  time-separated correlators should be evaluated with caution. The issue can be even more complicated for gauge theory where the Wilson line should be incorporated.

The second scheme exploits a connection between PDF and the hadronic tensor.   The hadronic tensor for the $d$-dimension theory is defined as correlations of currents $J^\mu$,
\begin{equation}
W^{\mu\nu}(q)=\text{Re}\int d^dx e^{iqy} \bras{P}T\{J^\mu(y)J^\nu(0)\}\kets{P},
\end{equation}
which can characterize the many-body wavefunction $\ket{P}$ through correlations and consequently reveal the hadronic structure to some degree.   The connection between PDF and $W^{\mu\nu}(q)$ is established as (via collinear factorization),
\begin{equation}
W^{\mu\nu}= \sum_{i} f_i \otimes \hat{W}_i^{\mu\nu}.
\end{equation}
Here $\otimes$ represents convolution,  $i$ stands for parton species and  $\hat{W}_i^{\mu\nu}$ is the partonic tensor. Thus, with $W^{\mu\nu}$, the parton distribution $f_i$ of species $i$ can be derived.  The nontrivial relation between PDF and hadronic tensor is indicated in the convolution.
%As both PDF and hadronic tensor can be used for deriving the cross section, it is reasonable to consider they can be equivalence in revealing the hadronic structure in some sense.

The remaining task is to evaluate $W^{\mu\nu}(q)$ on the state $\ket{P}$, which can refer to linear response method on a quantum computer.  Still, the preparation of the state $\ket{P}$ is quite involved, and Ref.~\cite{nu_20} points out that quantum adiabatic state preparation may be employed.

The above proposal suggests a way for studying parton physics on a quantum computer, but a demonstration is still awaited. This requires to efficiently prepare $\ket{P}$ (possible with VQE), evaluating the hadronic tensor, and extracting PDF through global fitting to the results of hadronic tensor obtained from quantum computing.  Another important question is which lattice Hamiltonian should be a good starting point.
The light-front Hamiltonian of QCD looks promising as it already have provided a framework for \textit{ab initio} calculation of proton structure. Remarkably, VQE based on the light-front Hamiltonian has just been demonstrated for calculating structure of pion~\cite{kreshchuk_light-front_2020}.
%Alternatively, large-momentum effective theory may be also a candidate  once the corresponding lattice Hamiltonian can be formulated.

\subsection{Quantum advantage: scattering in scalar quantum field theory.}
Scattering is central to nuclear physics as it is almost the only available experimental method.  Consequently, calculating scattering amplitude is what in theory needed to do for a comparison with experiments.  Different machinery of computational methods have been developed but there are some fundamental challenges. The difficulty can be revealed through a minimal model of quantum field theory, the $\phi^4$ theory, which is a scalar theory with quartic self-interactions.  When the coupling approaches the phase transition point, the perturbative method becomes unreliable; and even in the weak-coupling regime, precision can be not controllable with the Feynman diagram calculation. Also, it is beyond the ability of lattice field theory which is good at calculating static properties, while scattering is a time-evolution problem.  On the other hand, time-evolution can be implemented on a quantum computer and it is thus expected that scattering can be investigated efficiently.

A technical challenge is that the number of degree of freedom per unit volume is infinite,
and it is necessary to give a mapping that can assign qubits to physical degree of freedom, in a controllable way under desired accuracy.  For instance, truncation of Fock space can be applied, since the number of particles is constrained by the energy of incoming particles. Then, given the number of qubits, preparation of incoming two particles as two wave packets, and their scattering, should be formulated in the language of quantum circuit. The quantum advantage is that the depth of the quantum circuit scales only polynomial with the number of particles, energy and the desired precision, for both weak and strong couplings~\cite{jordan_12}.

%Physically, the scattering problem of $\phi^4$ can be characterized by energy of incoming particles and the coupling strength. With those as guild lines, Ref. works a scheme that proves quantum advantages for scattering problem in $\phi^4$ quantum field theory across all parameter regimes.

The space is discretized as a d-dimensional $L\times...\times L$ lattice $\Omega$ with lattice spacing $a$.  Let us start with the lattice Hamiltonian,
\begin{eqnarray}
H&=&\sum_{\bd{x}\in\Omega} a^d [\frac{1}{2}\pi(\bd{x})^2+\frac{1}{2}(\bigtriangledown_a\phi)^2(\bd{x}) \nonumber \\
&+& \frac{1}{2}m_0\phi(\bd{x})^2+\frac{\lambda_0}{4!}\phi(\bd{x})^4].
\end{eqnarray}
Here $\phi(\bd{x})$ is field operator on the site $\bd{x}$, $\pi(\bd{x})$ is the conjugate field satisfying $[\phi(\bd{x}),\pi(\bd{y})]=ia^{-d}\delta_{x,y}$, $\bigtriangledown_a$ is a finite-difference operator, $m_0$ is the particle mass for the non-interacting theory $H_0$ corresponding to $\lambda_0=0$.

The first issue is to represent the infinite dimensional field $\phi(\bd{x})$  in terms of qubits.
Setting $\phi_{max}$ as a cutoff and $\delta_\phi$ as increments, $n_b=O(\log_2(\phi_{max}/)\delta_\phi)$ qubits are required per site.  With a constraint of energy $E$,  the cutoff is $\phi_{max}=O(\sqrt{L^dE/a^dm_0^2\varepsilon})$ for desired fidelity $1-\varepsilon$ to the original state.  The energy constraint also introduce a cutoff $\pi(\bd{x})$ to $\pi_{max}=O(\sqrt{L^dE/a^d\varepsilon})$. Note that $\pi_{max}=1/(a^d\delta_\phi)$ as  $\phi(\bd{x})$ and $\pi(\bd{x})$ are conjugate. Then, it follows that $n_b=O(\log_2(L^dE/m_0\varepsilon))$.

Two incoming particles should be expressed as a many-body initial state with two separated wave packets on the lattice. The initial state can be prepared in an adiabatic way, which involves two steps: firstly,  two separated wave packets are produced on the vacuum of the non-interacting Hamiltonian $H_0$, where the vacuum is a Gaussian state and can be constructed exactly;  Secondly,  interaction is turned on and the system evolves adiabatically with Hamiltonian parameterized by $H(s(t))=s(t)H_0+(1-s(t))H$ , where $H(0)=H_0$, $H(1)=H$ and $s(t)$ is the adiabatic path.  Note that the wave packets will propagate and broaden since it is not an eigenstate of $H_0$, additional backward evolution is required to undo unnecessary dynamical phases. The time complexity for the initial state procedure scales as $O((1/\varepsilon)^{1+d/2})$. The system of two incoming particles then evolves for a period that scattering occurs. Then, the interaction is adiabatic turn-off and the scattering result is sampled by measuring the number operator of momentum modes defined in the free theory.  With scaling analysis via effective field theory, the algorithm shows a complexity polynomial with the $\varepsilon$, e.g., in 1D its $O((\frac{1}{\varepsilon})^{-1.5-o(1)})$.  Remarkably, at strong coupling, the time complexity still scales polynomial with $1/(\lambda_c-\lambda_0)$~($\lambda_c$ is the quantum phase transition point), the momentum of incoming particles $\bd{p}$, the maximum kinematically allowed number of outgoing particles $n_{out}$ and the precision $\varepsilon$. This is in contrast to the known classical algorithm scaling  exponentially with $1/(\lambda_c-\lambda_0)$ and $1/\varepsilon$,  thus showing an exponentially quantum speed-up.

The merit of this work is to make clear the point of solving quantum field theory on a quantum computer with quantum advantage. A scheme to reach this goal has also been pointed out. Still, some subroutines may be improved to make the algorithm more implementable in the NISQ era, and there indeed are some following jobs with more detailed algorithms~\cite{klco_19}.

\subsection{Non-equilibrium dynamics}
In physics we often need to study non-equilibrium quantum systems at finite temperature.  For nuclear physics,  quark-gluon plasma, which is produced in heavy-ion collision or in the expansion of the early Universe,  is an outstanding example but still limited knowledge is known, as simulation with classical methods are very hard.  Basically, studies of such quantum systems involves time evolution of density matrix. As a quantum computer essentially manipulate pure quantum states with unitary operation, an extension to density matrix needs additional treatment.  One can either refer to a subsystem by tracing out the ancillary, or view the density matrix as an ensemble of pure quantum states with a classical distribution.

Consider a density matrix $\rho_0$ as the initial state for a quantum system, then its real-time evolution under $H_1$ is given by $\rho(t)=e^{iH_1t}\rho_0e^{-iH_1t}$.  The initial state can be set as equilibrium state for the system under $H$ at inverse temperature $\beta=1/T$, namely quantum Gibbs state, $\rho_0\equiv\rho(\beta)=e^{-\beta H}$, which satisfying the symmetric Bloch equation,
\begin{equation}\label{eq:bloch}
\frac{d\rho(\beta)}{d\beta}=-\frac{1}{2}(H_0\rho+\rho H_0),
\end{equation}
with initial condition $\rho(0)=1$.
In Ref.~\cite{lamm_18}, a hybrid quantum-classical scheme has been developed for the real-time evolution of density matrix in the context of nuclear physics.  The initial state is obtained by solving Eq.~\eqref{eq:bloch} with density
matrix quantum Monte Carlo algorithm (DMQMC), leading to an approximation of $\tilde{\rho}_0\approx\rho_0$, with
$\tilde{\rho}_0=\sum_p\chi_p\ket{a_p}\bras{b_p}$.   Such an approximation
is completed on a classical computer with enough samples of $\ket{a_p}\bras{b_p}$. With the approximated initial state,  time-dependent observable
is evaluated as~(with a substitution $\tilde{\rho_0}\rightarrow\frac{1}{2}(\tilde{\rho_0}+\tilde{\rho_0}^*)$ ),
\begin{eqnarray}
&&\langle O(t)\rangle=\text{Tr}(Oe^{iH_1t}\rho_0e^{-iH_1t})/\text{Tr}\rho_0 \nonumber \\
&&\approx \frac{\sum_{p}\text{Tr}(Oe^{iH_1t}[\chi_p\ket{a_p}\bras{b_p}+\chi_p^*\ket{b_p}\bras{a_p}e^{-iH_1t}])}{2\text{Tr}\tilde{\rho_0}}. \nonumber \\
\end{eqnarray}
The task thus can be decomposed to evaluate observable $O$ for each component, including the diagonal term, e.g., $\bras{a_p}e^{iH_1t}Oe^{-iH_1t}\kets{a_p}$, and the off-diagonal term,  such as $\bras{a_p}e^{iH_1t}Oe^{-iH_1t}\kets{b_p}$. The former corresponds to perform measurement $O$ on a state evolved with $e^{-iH_1t}$ from initial state $\ket{a_p}$ or $\ket{b_p}$. And for the later, one can replace initial state with $\frac{1}{\sqrt{2}}(\ket{a_p}\pm \ket{b_p})$.

In total, the hybrid quantum-classical approach adopts a strategy that assign quantum resource for the real-time evolution part which is hard for the classical computer, while the classical part can utilize the state-of-art classical algorithm since it is good at solving steady states. A combination of both makes this approach promising on NISQ quantum processors for studying non-equilibrium dynamics.

\section{Outlook and summary}\label{sec:outlook}
At this stage, we have given an introductory review of quantum computing for nuclear physics, but a throughout investigation  is still beyond the scope of this  review, given recent rapid expanding of this field. We further briefly give an outlook that focuses on this field within the NISQ era. Although formulations of lattice gauge theory for quantum computing have been developed in a few different approaches, it is clear that there still is a considerable gap for simulating QCD on NISQ quantum processors with limited quantum resources, and it is anticipated that more efficient schemes of expressing degree of freedom of gauge field and its coupling to matter field would be developed. On the other hand, incorporating quantum algorithms as subroutines into a hybrid quantum-classical algorithm can be a very promising approach for solving nuclear physics with practical results.

Another observation is that there is still a lack of consensus on the goal; perhaps it is inspiring that a road-map can be proposed for demonstrating quantum advantages for nuclear physics on a quantum computer, in accordance with the scaling-up of near-term quantum processors and error mitigation techniques .  Moreover, software methodology is important, as it enables systematic developments and rapid innovations on both quantum algorithms and applications to nuclear physics. Compared with quantum computing for chemistry~\cite{McClean_2020} and quantum machine learning~\cite{bergholm_pennylane_2020}, open-source packages are still lacked for nuclear physics, and it awaits for such packages for making an attempt to study nuclear physics on a quantum computer more friendly.

In summary, we have given an brief review on recent advances on quantum computing for nuclear physics.
We have clarified two different approaches from analog and digital quantum simulator, and the review has focused on the latter. We have outlined how degrees of freedom of nuclear physics problems can be mapped into a formula of qubits that is solvable on a quantum computer, and the corresponding quantum algorithms for both static properties and real-time evolution have been shortly discussed. Concretely, we  have given some examples from recent outstanding works, ranging from simulation of lattice gauge theory, nuclear structure as well as quantum advantage in terms of time complexity for scattering problem. Lastly, we have pointed out that lots of efforts are still required to make quantum computing a playground for investigating nuclear physics on near-term quantum devices.

\begin{acknowledgments}This work was supported by  the Key-Area Research and Development Program
of GuangDong Province (Grant No. 2019B030330001), the National Key Research
and Development Program of China (Grant No. 2016YFA0301800), the National Natural Science Foundation of China (Grants No. 12005065, No.12022512, No.12035007, and No. 12074180), and the Key Project of Science and Technology of Guangzhou (Grants No. 201804020055 and No. 2019050001).
\end{acknowledgments}
%\begin{acknowledgments}
%This work was supported by the Key-Area Research and Development Program
%of GuangDong Province (Grant No. 2019B030330001), the National Key Research
%and Development Program of China (Grant No. 2016YFA0301800), the National Natural Science Foundation of China (Grants No. 91636218 and No. U1801661, Grants No. 12005065), the Key Project of Science and Technology of Guangzhou (Grant No. 201804020055).
%\end{acknowledgments}

%\section*{References}
%\bibliographystyle{iopart-num}
%\bibliography{nuclear}
%merlin.mbs apsrev4-1.bst 2010-07-25 4.21a (PWD, AO, DPC) hacked
%Control: key (0)
%Control: author (0) dotless jnrlst
%Control: editor formatted (1) identically to author
%Control: production of article title (0) allowed
%Control: page (1) range
%Control: year (0) verbatim
%Control: production of eprint (0) enabled
%

\end{document}